\documentclass[epj]{svjour}
\usepackage{graphicx}
\begin{document}
\title{Molecular Dynamics simulations of 
palladium cluster growth on flat and rough graphite surfaces\\}
\titlerunning{MD simulations of cluster growth}
\headnote{submitted to Eur. Phys. J. AP}
\authorrunning{P. Brault and G. Moebs}
\author{
Pascal Brault\inst{1,}\thanks{\email{Pascal.Brault@univ-orleans.fr}} \and
Guy Moebs\inst{2,}\thanks{\email{Guy.Moebs@crihan.fr}}
}
\institute{Groupe de Recherches sur l'Energ\'etique des Milieux Ionis\'es,
UMR6606 CNRS-Universit\'e d'Orl\'eans BP 6744, 45067 Orl\'eans Cedex 2, France
\and
Centre de Ressources Informatiques de Haute-Normandie, 745
 Avenue de l'Universit\'e, 76800 Saint Etienne du Rouvray, France}

\abstract{
Parallel Molecular Dynamics simulations are conducted for describing growth on
surfaces with different kind of roughness : a perfect ordered crystalline flat
 graphite surface, a disordered rough graphite surface and flat surface with 
 an ordered localized defect.  It is shown that disordered 
rough surfaces results in a first step to reduction of the sticking coefficient,
increased cluster density, size reduction. Structure of the clusters shows
the disappearance of the octaedral  site characteristic of compact structure.
 Isolated defect induces cluster-cluster interactions that modify 
growth compared to perfect flat surface. Kinetic study of growth shows power law
$t^{\alpha z}$ evolution for low  impinging atom kinetic energy. Increasing kinetic energy,
on all kinds of surfaces, results in a slightly larger exponent z, but fitting
by an exponential function is quite good too. Lattice expansion is favoured on
rough surfaces but increasing incoming atom kinetic energy weakens this effect.
}

\maketitle
\section{Introduction}
Cluster growth is a very important step in the early stages of thin film 
deposition \cite{vena84}. Modern methods are emerging for depositing atoms in 
high non equilibrium situation such as  pulse laser deposition \cite{basi02}, 
ion beam assisted deposition \cite{esch96}, plasma sputter deposition with high
ion to neutral flux ratios \cite{thom00,brau98,ross99}. The common feature of 
all these methods is
the ability to ensure deposition assisted by ion flux and with varying and 
controlled
 kinetic energy of the depositing species. Moreover, the ions impinging the 
 surface can be responsible for additional effects at the surface, i.e. creation 
 of defects, film densification. Ion or plasma 
 surface treatment can also be chosen for creating roughness 
expected to be suitable for thin film adhesion \cite{smit95}. Some theoretical 
attempts have been undergone for modelling thin film deposition due to the
availability of high performance computers \cite{gilm98,gilm00,wadl01}. In this
context, an interesting feature that we describe here is the
understanding of the role of ions and high kinetic energies have on the 
nucleation  and growth of thin film, especially using Molecular Dynamics (MD)
 simulations \cite{mull86,mull87,srol96a,srol96b,brau02,lued89,lewi96,lewi97}.
Main interest in MD is to carry precise calculations, provided that potential
interactions are known, and to give
deep insight into dynamical process \cite{shap98}. Especially cluster growth of 
transition metal element such as 
Pd, Pt, Rh, Ni, Cu, ... is subject to many studies of the potential due to the
interest in catalysis. So they fall within the capability of MD simulations 
because of the availability of now  good interactions potentials issued 
from Embedded Atom Method (EAM) \cite{daw84} or Tight Binding - Second Moment 
Approximation (TB-SMA) \cite{rosa89}. Both are true N-body potentials and are analytical. 
TB-SMA potentials are preferred on the basis of discussion in
Ref.~\cite{desj98,rosa89,legr90}.
This article is intended to examine how deposition conditions like kinetic
energy of incoming atoms, defects or roughness of the virgin surface play 
a role in the cluster growth. Such a study is expected to be suitable for
describing growth in plasma sputtering or pulsed laser
deposition. Indeed, calculating sticking coefficients, radial density functions and 
statistical informations issued from snapshots of the surface will give relevant
informations about initial steps of growth via clusters. So the present calculations
are intended to show how cluster growth is dependant on the nature of roughness
how kinetic energy of incoming atoms can minimize or not roughness effects.
For this purpose we simulate the growth of 
palladium on three kinds of surfaces : a perfect ordered crystalline flat
 graphite surface, a disordered rough graphite surface and flat surface with 
 an ordered localized defect.
The next section will describe both the MD algorithm and the interaction
potentials involved in the calculations. The results will be presented and 
discussed in the third section. The overall work will be summarized in the
concluding section.
\section{Molecular Dynamics}
Molecular Dynamics (MD) is a simulation technique in which classical equations
of motion are solved for a set of atoms or molecules. This leads to the
 well-known classical Newton set of equations describing the motion of atoms.
  This can  be written in the form :
\begin{equation}
m_i \frac{\partial^2 }{\partial t^2}\vec{r}_i = \displaystyle \sum_{\lambda}\vec{F}_{i}(\lambda)
\end{equation}
where $m_i$ is the mass of the $i^{th}$ incoming atom interacting through the 
the forces $\vec{F}_{i}(\lambda)$. $\lambda$ stands for surface atoms and adsorbed atoms .
 In principle we should have the same set of equations for the surface atoms: 
 they interact among themselves and also with adsorbed atoms.  In the following, 
 the surface atoms remain at their initially fixed posincluding particle interactions with surface
   atomsitions. This is justified 
here for two reasons : first, graphite lattice is a very rigid lattice, second, 
impinging atom energies are well below graphite carbon displacement energy. 
The disordered 
 surface is obtained by randomly displacing the atoms from their known 
 equilibrium sites. When substrate atoms do not move,
  it is necessary to find a way  for dissipating energy through the solid 
  for allowing bonding to the surface. As a first attempt, we make use of 
  quenched molecular dynamics \cite{mott92}: if  
 at a time step $\vec{F}.\vec{v} < 0$,($\vec{F}$ is the total force exerted
 on the considered atom), then the velocity $\vec{v}$ of the atoms
  is reset to  a velocity randomly chosen in a velocity Maxwell distribution 
  at surface temperature which is fixed here to T$_s$ = 300 K (in Ref.~
  \cite{mott92}, velocities are reset to 0). This is justified because the 
energy transfer considered here  will not affect the graphite lattice due to 
its high stiffness. In that case, diffusion remains allowed and is random.

Simulating deposition needs to release atoms one after each other with a time 
delay $\Delta t$ suitable for comparison with experiments, i.e. either the depositing
atom flux reproduces exactly the one encountered in experiments or 
this time delay $\Delta t$ is
 sufficient for allowing thermal relaxation of the already deposited atoms 
 and/or thermal relaxation of surface atoms (when they are allowed to move). 
 In our case we choose the latter asumption, which allows us to reduce computer 
 time, even if it leads to  large flux: this could be reasonable in sputter or
 pulsed laser deposition.
 $\Delta t$ is thus fixed to 2 ps. Increasing this time delay does not 
 change our results, which renders our assumption convenient.\\
Implementing suitable interatomic potentials is certainly the most important 
issue in molecular dynamics calculations. For describing transition metals 
like palladium, we use tight binding potential in the second moment 
approximation (TB-SMA)\cite{rosa89}. Such a potential is non pairwise in the 
sense that if atom $i$ interact with atom $j$, the atoms surrounding atom j are explicitly 
taken into account.The TB-SMA force equation acting on atom i due to atom j
surrounded by atoms k, can be written as :
\begin{eqnarray}
\vec{F}_{i} ( \mbox{Pd} -  \mbox{Pd} ) =
 \displaystyle \sum_{j \neq i, r_{ij} < r_c^{TB}} \left\{
 2 Ap \exp \left[ -p \left( \frac{r_{ij}}{r_0} - 1 \right) \right] \right.
 \nonumber\\
\left.- \frac{\xi q}{r_0} \left[ \frac{1}{\sqrt{ \mbox{E}_{i}^{b} } } +  
 \frac{1}{\sqrt{ \mbox{E}_{j}^{b} } }\right]
\exp \left[ -2q \left(  \frac{r_{ij}}{r_0} - 1 \right) \right] \right\} 
\frac{ \vec{r}_{ij} }{ r_{ij} } 
\end{eqnarray}
with

\begin{equation}
\mbox{E}_{i}^{b}  = \sum_{j \neq i} \exp \left\{ -2q \left( 
 \frac{r_{ij}}{r_0} - 1 \right) \right\} 
\end{equation}

and
\begin{equation}
\mbox{E}_{j}^{b} = \sum_{k \neq j} \exp \left\{ -2q \left( 
 \frac{r_{jk}}{r_0} - 1 \right) \right\} \nonumber
\end{equation}
where  $r_0$ is the first neighbour distance. For palladium $r_0$ = 0.275 nm. 
The interaction is cut off at $r_c^{TB}$ = 2.5$r_0$ (which includes neighbours
up to the 5$^{th}$).$\quad  r_{ij}$ is the interatomic distance 
between $i$ and $j$ atoms. $A,p,q,\xi$ are the TB-SMA parameters \cite{rosa89}.
Even if this potential looks like a two-body  form, it is needed for each $j$ atom 
to search for all neighbours within the cutoff radius $r_c^{TB}$ and to 
calculate the sum $E_j^b$, so non pairwise nature of the interactions become clear. 
This makes the calculations computer time consuming, especially when 
$r_c^{TB}$ becomes quite large. While $r_c^{TB}$ can be restricted to
$r_0$ in bulk materials 
(because bulk atoms only oscillate at their equilibrium position), when 
deposition 
simulations are conducted, it is necessary to use larger cutoff radii, 
especially for accounting interactions with diffusing atoms. The value we 
choose is the smallest which does not change the results. It allows taking 
into account 92 neighbours (for palladium), each neighbour interacting with 
its own 92 neighbours (when comparing to bulk materials). For ultrathin films,
 the number of neighbours can only be reduced at initial steps. Thus it becomes
  clear that high performance (parallel) computers are required for treating
long time deposition using such kind of interactions, which allows to treat 
a few thousands of incoming interacting particles with a few ten of thousand 
substrate atoms.

For interactions with fixed C atoms, we used a Lennard-Jones (LJ)
 12 - 6 potential:
\begin{equation}
\vec{F}_{i} ( \mbox{Pd} -  \mbox{C} ) = 24 ~\varepsilon_{ \mbox{\tiny Pd} - \mbox{\tiny C} }~
\sum_{j} \left\{ ~2 \left[ \frac{ \sigma_{ \mbox{\tiny Pd} - \mbox{\tiny C} } }
{ r_{ij} } \right]^{12} -
                    \left[ \frac{ \sigma_{ \mbox{\tiny Pd} - \mbox{\tiny C} } }
{ r_{ij} } \right]^{6} \right\}
\frac{ \vec{r}_{ij} }{ r^2_{ij} }
\end{equation}
This Pd-C  Lennard-Jones interaction potential is obtained by using the 
Lorentz-Berthelot mixing rule \cite{lued89,liem94,wu96}:
$\varepsilon_{ \mbox{\tiny Pd} - \mbox{\tiny C} }= \left(
 \varepsilon_{ \mbox{\tiny Pd}}\varepsilon_{ \mbox{\tiny C}} \right )
^{\frac{1}{2}}$ and $\sigma_{ \mbox{\tiny Pd} - \mbox{\tiny C} } = 
\frac{ \sigma_{ \mbox{\tiny Pd}}+\varepsilon_{ \mbox{\tiny C}} }{2}$.
 The LJ Pd-Pd interaction parameters are:$\quad \varepsilon_{ \mbox{\tiny Pd}}$ 
 = 0.426 eV and $\sigma_{\mbox{\tiny Pd}}$ = 0.252 nm \cite{hali75}. 
 The parameters for C-C interactions can be found in reference \cite{stee73} 
and the values are: $\varepsilon_{ \mbox{\tiny C}}$ = 2.414  10$^{-3}$ eV 
and  $\sigma_{\mbox{\tiny C}}$= 0.340 nm. This gives: 
$\varepsilon_{ \mbox{\tiny Pd} - \mbox{\tiny C} }$= 0.02627 eV and $\sigma_{
\mbox{\tiny Pd} - \mbox{\tiny C} } $ = 0.299 nm \\
The equations of motion are solved using the Verlet velocity algorithm
\cite{swop82,alle87}. 
A link-cell list is used to speed-up the computations in conjunction with Verlet
lists for which radius $r_v = 2.7 r_0$. Due to the non-pairwise
 interactions, the CPU time is not reduced to O(N), where N is the number of 
 particles. The time step required for numerical integration of equations of 
motion is dt = 1 fs. It was verified it is enough up to impinging energy of 2 eV.
For parallel implementation, we used the atom-decomposition scheme also known 
as replicated data method \cite{plim96}. This method is used because the filling
 of space is expected not to be complete so spatial-decomposition scheme can not 
 be an efficient scheme \cite{plim96}. The parallel instruction library OpenMP 
 has been found to 
be more efficient than message passing libraries as MPI because the calculations
 are performed on a shared memory supercomputer (SGI Origin 2000, 64 nodes,
 at CRIHAN, France). Briefly, N/p particles are treated by each p nodes during all the 
 course of the simulation. At the same time, all information about each particle
  is known from each p nodes. Here p is varied from 4 to 14 depending on the
  load, so parallel efficiency is maintained above 90 \%.

\begin{figure*}[h]
\begin{center}
  \includegraphics*{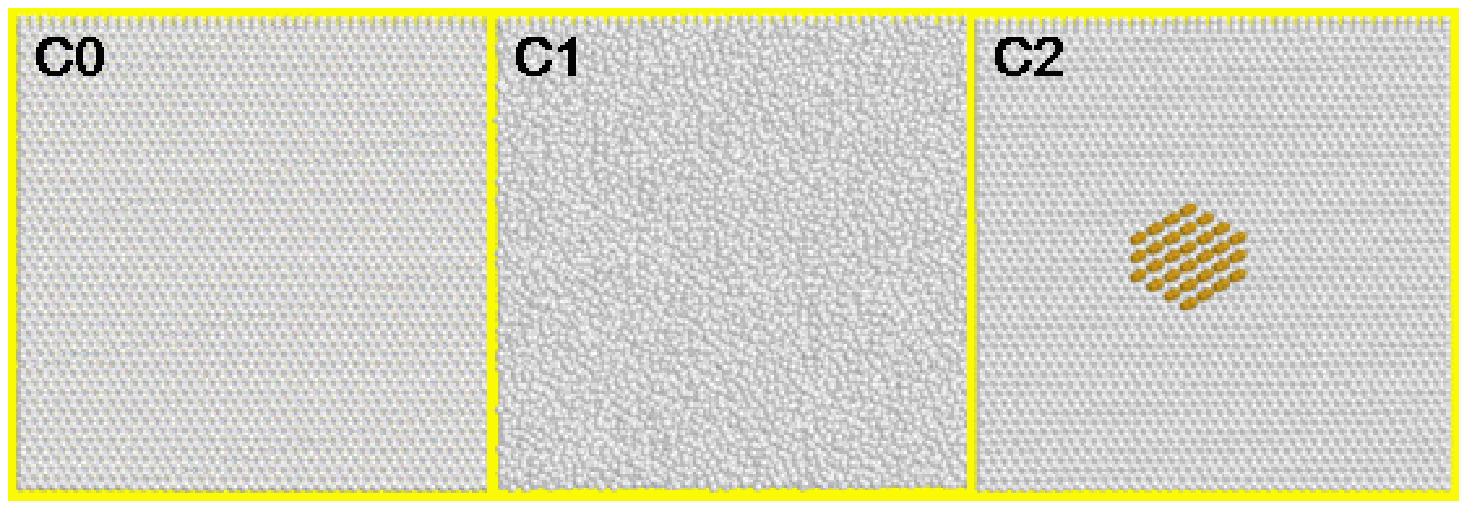} 
\end{center}
\caption{\label{fig:1} Simulated graphite surfaces (10.3 nm x 10.2 nm): 
C0, atomically flat; C1, with randomly disordered first outer plane;
 C2, flat with a localized defect. The defect is dark gray}
\end{figure*}

The simulated substrates are built from 3 atomic layers of 10.3 nm x 10.2 nm
leading to rigid substrates of 12369 C atoms. This is sufficient for taking into
 account all interactions between incoming Pd atoms and substrate carbon atoms. 
 In the following, one monolayer
  (ML) corresponds to 1521 atoms as for a
  Pd fcc (111) structure. Three kinds of surfaces have been built: 
  an  atomically smooth (surface C0); a highly roughened surface where atoms of the
  top layer are 
  randomly displaced in-plane by 50\% of the C-C distance, $a_{C-C}$ = 0.142 nm ,
  and  50\% of the interlayer distance, $h_{C}$ = 0.335 nm, perpendicular
  displacement (surface C1); 
  and a flat surface with a localized defect of size $\approx$ 2nm (surface C2).
  This defect is composed of 60 carbon atoms 0.2 nm vertically elevated above
  their natural site. This latter structure is typical from ion or plasma 
  irradiated surface \cite{hahn99,rous02}.
  These three surfaces are displayed in Fig.~\ref{fig:1}.
  
  Each palladium atom are randomly launched every $\Delta t$ = 2 ps above the
  surface. There are no interactions between Pd in the gas phase. All the Pd atoms 
  interact simultaneously with other palladium atoms within the cutoff  radius 
  $r_c^{TB}$ and  with carbon rigid substrate atoms with a cutoff  radius $r_c 
= 2.5\sigma_{\mbox{\tiny Pd-C}}$. The   calculations have been performed at 0.31 ML
 (500 Pd atoms), 0.62 ML 
  (1000 Pd atoms) and 0.93 ML (1500 Pd atoms), 1.08 ML (1750 atoms), 1.24 ML
  (2000 atoms), 1.40 ML (2250 atoms), 1.55 ML (2500 atoms), 1.71 ML (2750 atoms) 
  and 1.86 ML (3000 atoms). Each set of calculation last for 6 ns (= 3000 atoms
  x 2 ps), for calculating interactions between 3000 incoming atoms among
  themselves and with 12369 substrate atoms. All sets of calculations use the
same random number sequence. This means that differences between calculations 
for different surfaces or different energies will only originate from specific interactions.
   
    The Pd initial mean kinetic 
  energies are 0.026, 0.1 and 1. eV. In the former case, it simulates a metal
  vapor at 
  T$_g$ = 300 K. This occurs when sputtered Pd atoms travel across a buffer gas 
  at sufficient pressure. When T$_g$ = 0.1 eV, it is consistent with resistive or
  e-beam evaporation, which produces a vapor at the vaporisation temperature,
  which just lies in the range 0.1 eV.  The latter case simulates a vapor at 
   temperature T$_g$ = 1.0 eV. This occurs for sputtering experiments at low 
pressure where a small amount of buffered gas only randomizes sputtered atom 
velocities, without not too much energy loss ($\approx$ 50\%). Peak energy of the
sputtered atom energy distribution is half vaporisation energy in vacuum
( sputtered atom Thompson energy distribution). Then in all cases, initial 
velocities are chosen in Maxwell-Boltzmann distribution at the given 
temperature T$_g$ = 0.026, 0.1 and 1.0 eV, with random corresponding incident
angles.
   
   For each surfaces C0, C1, C2 surfaces, three sets of deposition simulations
   are run corresponding to T$_g$ = 0.026, 0.1 and 1.0 eV. Each run lasts 6 ns 
real time displaying 3000h CPU time shared by 4 to 14 processors on SGI 
Origin 2000 supercomputer (ILLIAC8 at CRIHAN). Altogether, calculations last 
27000 hours corresponding to about 3 years on an equivalent single processor 
computer.
\section{Results and discussion}
\begin{figure}[h]
\begin{center}
  \includegraphics*[width=\linewidth,scale=1.2]{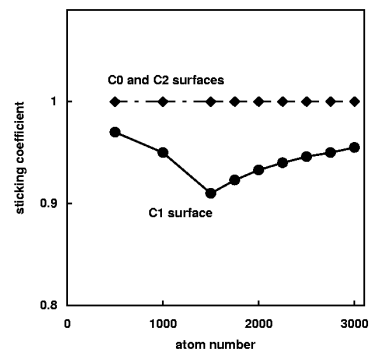}  
 \end{center}
\caption{\label{fig:2} Evolution of the sticking coefficient on surfaces 
C0, C1 and C2. Note it is independant from impinging kinetic energy}
\end{figure}

The sticking coefficient is the ratio of the incoming atom number to the adsorbed
atom number. It informs about adsorption processes \cite{brau97}.
Fig.~\ref{fig:2}. displays the evolution of the sticking coefficient for all 
conditions.
The sticking is always unity except for surface C1, where it first decreases
and then further increases. This behaviour is independant from the explored
 range of kinetic energy. Due to atomically sized roughness of surface C1, at
low coverage, some atoms ($\approx$ 5-10\%) may not find a stable adsorption 
site and so desorbs. Then,
sticking coefficient first decreases. When clusters are already formed and are
stable  enough, they offer a large capture area where sticking is unity. Thus 
it allows now to increase the
sticking coefficient towards unity. That is the meaning of the initial
decrease of sticking as plotted in Fig.~\ref{fig:2}.

\begin{figure}[h]
\begin{center}
  \includegraphics*[width=0.9\linewidth]{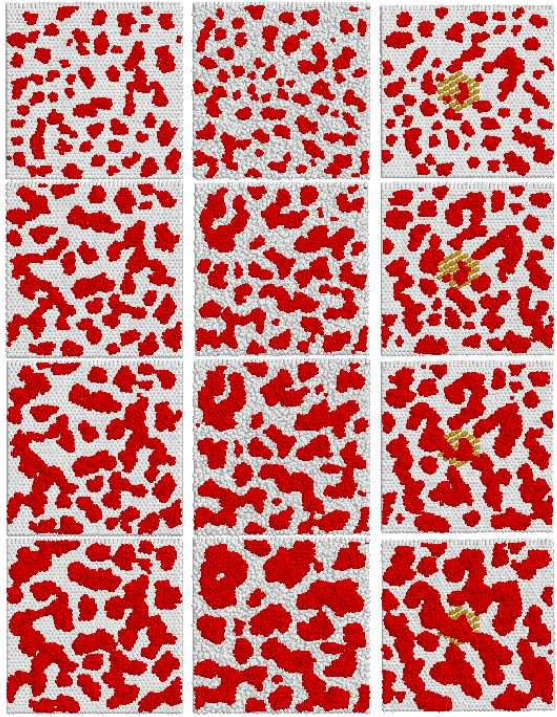}  
 \end{center}
\caption{\label{fig:3} Snapshots of Pd clusters deposited on surfaces C0 (left
column), C1 (middle column) and C2 (right column). Size of imulated surfaces is
 10.3 nm x 10.2 nm. The mean kinetic energy is $E_c$ = 0.026 eV. From top to bottom 
the number of deposited atoms is respectively 1000, 1500, 2000 and 3000.}
\end{figure}

\begin{figure}[h]
\begin{center}
  \includegraphics*[width=0.9\linewidth]{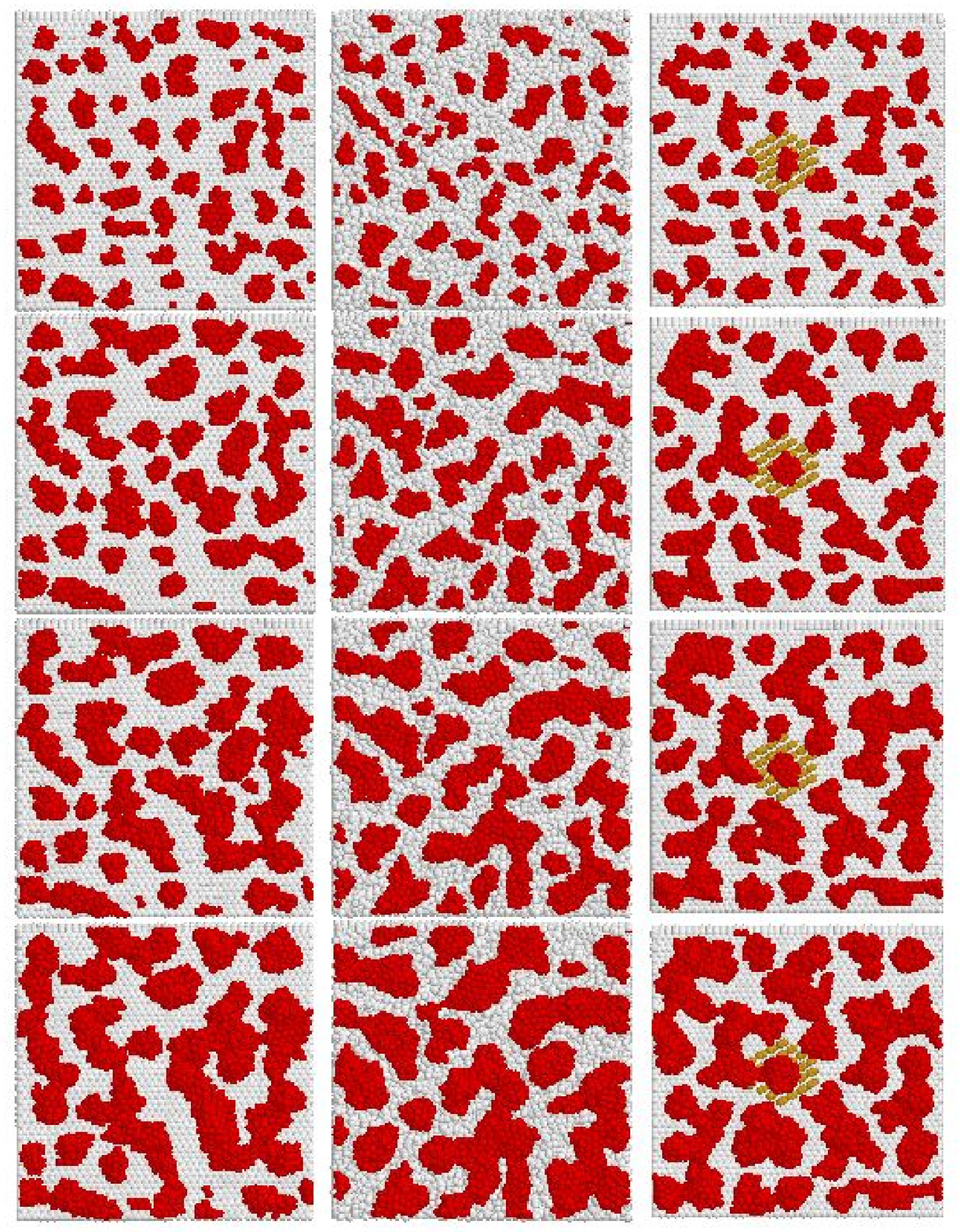}  
 \end{center}
\caption{\label{fig:4} same as Fig.~\ref{fig:3} but for  $E_c$ = 0.1 eV}
\end{figure}

\begin{figure}[h]
\begin{center}
  \includegraphics*[width=0.9\linewidth]{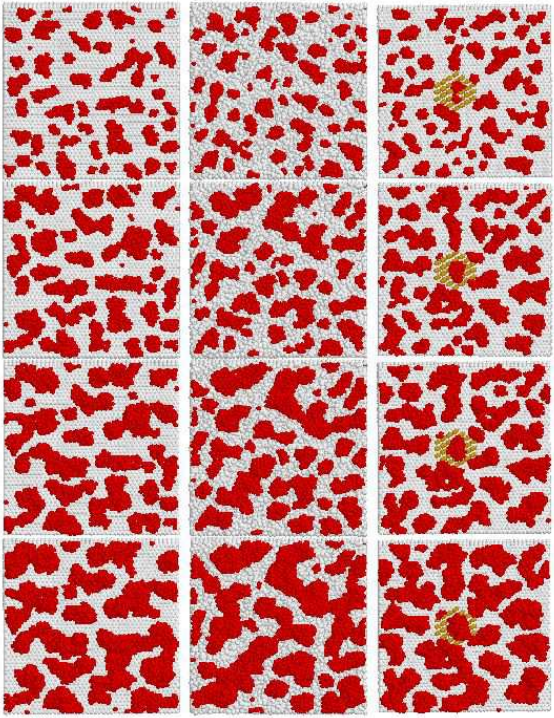}  
 \end{center}
\caption{\label{fig:5} same as Fig.~\ref{fig:3} but for  $E_c$ = 1.0 eV}
\end{figure}

\begin{figure}[h]
\begin{center}
  \includegraphics*[width=\linewidth]{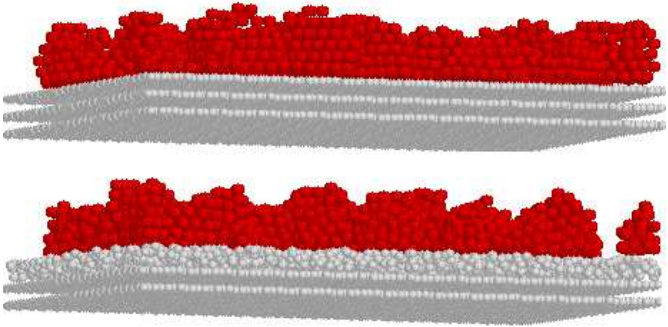}  
 \end{center}
\caption{\label{fig:6} Edges of cluster on surfaces C0 (top) and C1 (bottom)}
\end{figure}

\begin{figure}[h]
\begin{center}
  \includegraphics*[width=\linewidth]{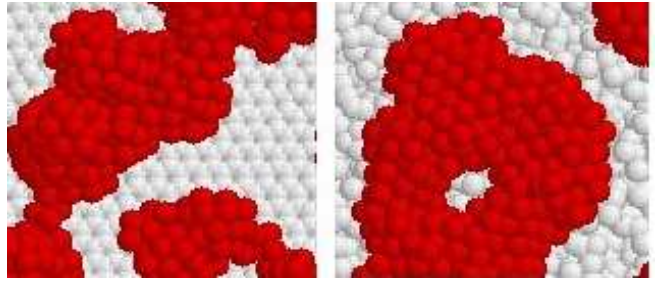}  
 \end{center}
\caption{\label{fig:7} Left picture is the surface of a cluster on flat
graphite C0. The cluster outer surface is rough. On right picture, cluster on 
rough graphite C1, exhibits flatter surface}
\end{figure}

On Fig~\ref{fig:3}-\ref{fig:5}  are plotted significant
 snapshots for the three surfaces at the three energies studied. 
In all cases, growth exhibit clusters
which are more or less meandering. The cluster mean height is around 6 atom
diameters (i.e. around 1.3 nm for Pd atoms) for 3000 deposited atoms
(see Fig.~\ref{fig:6}).
Looking at the shape of the clusters, we observe on Fig.~\ref{fig:7} that clusters are 
more flat on rough surface C1 than on flat C0 and C2. In fact these two situations are
 reversed: clusters have rough top and flat bottom on the surfaces C0 and C2 compared 
to clusters that have flat top and rough
bottom on the rough surface C1. This holds only for not too high clusters, i.e. when
memory of growth starting conditions is kept. Moreover
C1 clusters appear less ordered when looking at the edges (see Fig.~\ref{fig:6}).
On a flat surface, first atomic layers in the clusters are parallel to the surface plane.
On the contrary, clusters can not have the first layer parallel to a rough surface.

When looking at the same surface C0, C1 or C2 and examining cluster shapes
at a fixed coverage, clusters become more compact (meandering has now a lower
extent) as impinging atom kinetic energy increases. For example, surface C0
(left column of Fig.~\ref{fig:3}-\ref{fig:5}) for $E_c$ = 0.026 eV has a more developed
meandering structure than for $E_c$=1.0 eV. Moreover, for a 3000 incoming atom 
number the shape is quite different for the three energies while the cluster
number remains the same (10 clusters when taking into account the cell 
periodicity). At fixed energy, and comparing among the three surfaces,
differences appear essentially at high coverage.
For low energy (Fig.~\ref{fig:3}), compacity is favoured on the rough surface. At
highest energy no clear behaviour can be drawn.

Attention should also be paid to deposition on surface C2. Surface C2 differs
from C0 only by the 60 elevated (0.2 nm) surface carbon atoms in its center.
Simulations are done with the same random number sequence in each set of
calculations. This allows direct comparison between results. Indeed, differences
only originate from interactions due to the kind of surface. It is interesting
to observe differences between surface C0 and C2 which come from the isolated
defect. All the clusters on the whole surface are affected by the defect. Only
pre-existing cluster on the defect (center of Fig.~\ref{fig:3}) is growing and
enlarging on the defect. Other clusters remain at the edge of the defect. For
the 1000 incoming atoms on Fig.~\ref{fig:3}, a cluster is disrupted by the edge 
of the
defect. On C0, a larger cluster exists below the center of the simulation cell.
On C2, one part on the cluster remains on the defect, the other one being located
along the edge of the defect. For 3000 atoms deposited, small clusters
are finally coalescing, bridging the defect. But the same long cluster does 
not exists on C0. This means that cluster-cluster interactions are also driven 
by the isolated defect. Highest kinetic energy does not overcome this effect.
This can be due by a too high kinetic barrier at the cluster edges i.e. diffusing atoms
 do not gain enough kinetic energy to climb on or over the defect. This is 
consistent with the experimental observation of step decoration on graphite.
Fig.~\ref{fig:5} also displays the same kind of differences in growth shapes 
between C0 and C1. 

Informations that can be extracted from snapshots are cluster densities $N_c$, 
cluster mean sizes $\bar{d}$. The evolution
against deposition time or equivalently impinging atom number N give information
about growth modes \cite{vena84,zink92,meak93,viov88}.

Low coverage luster densities $N_c$  are larger on the rough surface at the lowest
kinetic energy. Increasing kinetic 
energy minimizes the effect of roughness on initial cluster density as can be
seen in Table~\ref{tab:1}.
\begin{table}[h]
\caption{\label{tab:1} Low coverage cluster densities $N_c$ on surface C0, C1 and C2 for
500 deposited atoms}
\begin{tabular}{cccc}
\hline\noalign{\smallskip}
$E_c $ & $N_c$ on C0 & $N_c$ on C1 & $N_c$ on C2 \\
 (eV)  &  (10$^{12}$cm$^{-2}$) &(10$^{12}$cm$^{-2}$) &(10$^{12}$cm$^{-2}$)  \\
\hline\noalign{\smallskip}
0.026 & 64 & 73 & 65 \\
0.10 & 65 & 69 & 69 \\
1.00 & 61 & 69 & 65 \\
\hline\noalign{\smallskip}
\end{tabular}
\end{table}
When increasing deposition time the cluster densities are reduced and become
independant of the roughness.

The evolution of the cluster mean size is often addressed in term of power laws.
The most common growth power law is given by $\bar{d} \propto N^z$ for isolated
clusters and   $\bar{d} \propto N^{\alpha z}$ after cluster coalescence
\cite{viov88,fami89,beys90}, N being the incoming atom number.$z$ and $\alpha z$
are known as growth exponents and are deduced from statistical physics of growth 
phenomena. They were introduced for displaying common features of various phenomena
in term of universal scaling laws. For addressing
 this question, we gather the mean
areas $\bar{s}$ of the clusters at each energy and for each surface against all
incoming atom numbers in Table~\ref{tab:2}. Note this area is the projected area
perpendicular to the surface, thus it does not take into account the cluster
height. This is preferred because this is the same size that is deduced from 
electron microscopy pictures \cite{brau98}.
\begin{table*}[h]
\caption{\label{tab:2} Cluster area $\bar{s}$ (nm$^2$) evolution against incoming
atom number N for the three surfaces C0, C1,C2 and the mean kinetic energies
0.026, 0.1, 1.0 eV. Two kinds of fit are displayed : $\bar{s} \propto N^{2\alpha z}$
 and $\bar{s} \propto e^{\mu N}$. Values between brackets give a less good fit
 than power law does.}
\begin{tabular}{c|ccc|ccc|ccc}
\hline\noalign{\smallskip}
surface & &  C0& & & C1& & & C2 & \\
  & & $\bar{s}$ (nm$^2$)& & & $\bar{s}$ (nm$^2$)& & &$\bar{s}$ (nm$^2$) & \\
\hline\noalign{\smallskip}
N &0.026&0.10&1.00&0.026&0.10&1.00&0.026&0.10&1.00\\
\hline\noalign{\smallskip}
500  & 0.31 & 0.31 & 0.34 & 0.26 & 0.28 & 0.28 & 0.31 & 0.30 & 0.31 \\
1000 & 0.68 & 0.68 & 0.70 & 0.57 & 0.59 & 0.66 & 0.68 & 0.75 & 0.99 \\
1500 & 1.64 & 1.46 & 1.40 & 1.77 & 0.95 & 1.27 & 1.62 & 1.56 & 1.90 \\
1750 & 2.15 & 1.61 & 2.07 & 1.77 & 1.80 & 1.67 & 2.23 & 2.09 & 2.19 \\
2000 & 2.91 & 2.13 & 2.93 & 2.19 & 2.13 & 1.81 & 3.03 & 2.49 & 2.54 \\
2250 & 3.39 & 2.83 & 3.59 & 2.51 & 2.46 & 2.05 & 3.28 & 3.24 & 3.54 \\
2500 & 3.28 & 3.42 & 4.92 & 3.03 & 3.49 & 2.56 & 4.20 & 3.40 & 3.89 \\
2750 & 3.84 & 4.38 & 6.17 & 3.17 & 4.39 & 3.53 & 4.30 & 4.51 & 4.32 \\
3000 & 5.09 & 5.75 & 5.71 & 3.96 & 5.92 & 5.41 & 6.35 & 4.41 & 4.82 \\
\hline\noalign{\smallskip}
2$\alpha z$   & 1.6  & 1.6  & 1.75 & 1.5  & 1.7  &  1.5 & 1.6  & 1.6  & 1.5  \\
$\mu$ & (1.) & 1.1 & 1.3 & (1.) & 1.2 & 1.1 & (1.) & (1.) & (1.) \\
\hline\noalign{\smallskip}
\end{tabular}
\end{table*}

The evolution  of the equivalent size
$\bar{d}$ is recovered by $\bar{d} \propto \sqrt{\bar{s}}$ with  $ \bar{s}
\propto N^{2\alpha z}$. We obtained values in the range 
$\alpha$ z = 0.75 - 0.87 in agreement with
Beysens et al \cite{beys90} and Viovy et al \cite{viov88} (In this case,
$\alpha$ = 3 \cite{viov88}). This is consistent with static coalescence, i.e.
 clusters grow without moving until contact. It was 
also suggested that cluster growth could follow
exponential law \cite{jens94}. This is expected to occur when cluster diffusion
takes place. Thus we also evaluate an exponential fit : $ \bar{s}
\propto e^{\mu N}$ in Table~\ref{tab:2} with $\mu$ = 1.- 1.3. This means
that some diffusion competes with static coalescence.

On surfaces C0 and C2, increasing energy results in larger cluster area
at high coverage ($>$ 1.5 ML) : increasing kinetic energy enhances the atom
diffusion and diffusing atoms stick preferentially on the edges of large
existing clusters even if they are far away. This effect also occurs because Pd
cohesive energy is very high \cite{vena84}.

On rough surface, at high coverage, the largest areas are obtained for
intermediate energy 0.1 eV. At the same time, lower cluster areas/sizes are
obtained due to roughness which prevents sticking on too far cluster edges.

Power law behaviour (Table~\ref{tab:2}) confirms such analysis. Larger exponents 
(i.e.
fastest growth kinetics) are obtained for higher energies on C0 and 2$\alpha z $= 1.75.
On C2 there is no dependence against kinetic energy, which again shows the
special character of the C2 surface. For C1 it is obtained for the intermediate
energy too and 2$\alpha z $= 1.7.  The meaning of this behaviour is not yet clear. One 
could expect that increasing energy above a threshold defined by the roughness
scale first results in a behaviour similar to flat surface. But increasing again
kinetic energy, result in hampered motion due to multiple interactions with
atoms of the inhomogeneous surface. This would prevent further diffusion contrary
to a flat surface. And cluster area/size decreases consequently. When considering 
exponential fit, inspection of snapshots from
Fig.~\ref{fig:3}-\ref{fig:5} shows that cluster deformation operates as 
cluster diffusion. Additional information is required for more precise statements
about this question.
\begin{figure}[h]
\begin{center}
  \includegraphics*[width=\linewidth]{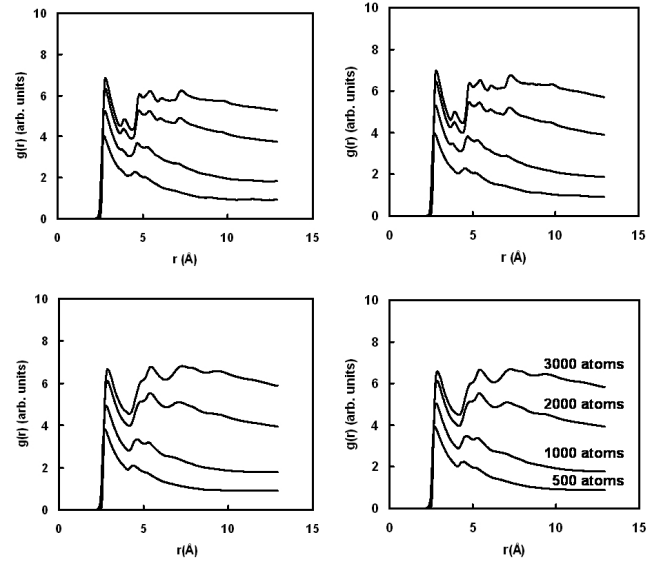}  
 \end{center}
\caption{\label{fig:8} Radial distribution functions for surfaces C0 and C1 
at two energies.}
\end{figure}

In Fig.~\ref{fig:8} are plotted radial (or pair) distribution functions 
$g(r) = \frac{V}{N^2}*\displaystyle\sum_i \displaystyle\sum_{j\neq i} 
\delta(\vec{r}-\vec{r_{ij}})$ \cite{alle87} for surface C0 and C1 at 
$E_c$ = 0.026 and 1.00 eV
This provides structural information about clusters, especially neighbour
distances, ordering, ...(this is quite close to the information given by 
experimental EXAFS (Extended X-Ray Absorption Fine Structure)). Direct inspection
of Fig.~\ref{fig:8} shows that g(r) are composed of several peaks. These peaks give
neighbour interatomic distances. Those corresponding to C0 are well defined and
becomes narrower and numerous when increasing the number of deposited atoms.
This means that long range order is taking place. This was already suggested by
the fcc (111) good stacking displayed on the upper panel of Fig.~\ref{fig:6}.figure
For C1, at all energies, peaks are broadened compared to C0, even for larger
clusters. 
\begin{table*}[h]
\caption{\label{tab:3} First $r_n$ and second $r_{nn}$ neighbour distance 
for all deposited atom numbers N and for surface C0 and C1. }
\begin{tabular}{c|ccc|ccc|ccc}
\hline\noalign{\smallskip}
surface & &  C0& & & C0 & & & C1  &  \\
  & &  $E_c$ (eV)& & & $E_c$ (eV)&  & &  $E_c$ (eV)&  \\
N &0.026&0.10&1.00&0.026&0.10&1.00&0.026&0.10&1.00\\
\hline\noalign{\smallskip}
  & & $r_n$ (nm)& & & $r_{nn}$ (nm)& & & $r_n$ (nm)&\\
\hline\noalign{\smallskip}
500  & 0.2688 & 0.2710 & 0.2710 & 0.3650 & 0.3661 & -      & 0.2710 & 0.2715 & 0.2721 \\
1000 & 0.2754 & 0.2743 & 0.2754 & 0.3759 & 0.3759 & 0.3792 & 0.2765 & 0.2798 & 0.2787 \\
1500 & 0.2787 & 0.2787 & 0.2776 & 0.3825 & 0.3857 & 0.3836 & 0.2819 & 0.2819 & 0.2841 \\
1750 & 0.2798 & 0.2776 & 0.2787 & 0.3857 & 0.3868 & 0.3857 & 0.2819 & 0.2841 & 0.2852 \\
2000 & 0.2809 & 0.2808 & 0.2798 & 0.3868 & 0.3879 & 0.3901 & 0.2852 & 0.2852 & 0.2852 \\
2250 & 0.2798 & 0.2808 & 0.2819 & 0.3879 & 0.3890 & 0.3920 & 0.2874 & 0.2852 & 0.2852 \\
2500 & 0.2808 & 0.2808 & 0.2808 & 0.3912 & 0.3901 & 0.3901 & 0.2885 & 0.2874 & 0.2874 \\
2750 & 0.2808 & 0.2808 & 0.2808 & 0.3901 & 0.3912 & 0.3901 & 0.2885 & 0.2874 & 0.2852 \\
3000 & 0.2808 & 0.2808 & 0.2808 & 0.3923 & 0.3912 & 0.3901 & 0.2885 & 0.2874 & 0.2852 \\
\hline\noalign{\smallskip}
\end{tabular}
\end{table*}
This means that ordering is not yet well established.
This is in agreement with the observation of Fig.~\ref{fig:6} where clusters appear
without any ordered stacking, contrary to C0. Moreover, the second and 
fifth neighbour peaks have disappeared, even for the highest deposited atom
number. This means that the octaedral site characteristic of compact structure has
disappeared. Recall that $2^{nd}$ neighbours of an atom in a fcc (111) plane are 
located in upper (3 neighbours) and lower (3 neighbours) (111) planes. 
The in-plane disorder persists up to six atomic distances
(which is the height of the clusters for 3000 deposited atoms ).
In Table~\ref{tab:3} is reported the evolution of  1$^{st}$ $r_n$ and  
2$^{nd}$ $r_{nn}$
neighbour distances for C0 surface and  1$^{st}$ $r_n$ neighbour distance for C1. 
Recall that for fcc structures, $r_n = \frac{a_0}{\sqrt{2}}$ and  $r_{nn} =
a_0$, $a_0$ being the lattice parameter ($a_0$ = 0.389 nm for Pd).
For bulk palladium $r_n$ = 0.275 nm and $r_{nn} = a_0$ = 0.389 nm.
Table~\ref{tab:3} shows for both C0 and C1 surface departure from these values.
At low coverage Pd atom find sites above the center of hexagonal ring of
graphite. Such hexagons are separated by a distance of 0.246 nm (lattice
parameter of graphite), thus at very low coverage when clusters are made 
of pair of  atoms, $r_n$ is closer to  0.246 nm rather than 0.275 nm. 
When increasing
atom number, clusters grow and increase their closest atom distance. They
crossed the ideal bulk value and a lattice expansion start as usual for
nanometer sized 
clusters. For palladium deposition on MgO and for cluster size less than 2.5 
nm lattice expansion as high as 0.8 \% was found experimentally using 
grazing incidence small angle X-ray scattering \cite{forn96}.  
For the C0 surface, at low coverage $r_n$ and $r_{nn}$ contraction is followed 
by further expansion. $r_n$  is in the range 2.0\%, while $r_{nn} = a_0$
expansion is only 0.5 \%.  The $r_n$ value is fairly high, but cluster lowest
size (cluster cross sectional size) is in the range 1.3 nm. For such low
size, lattice behaviour could be very distorded. For C1, due to roughness
induced mismatch, the $r_n$ expansion is increased up to 4.9 \%. But increasing
energy reduces $r_n$ expansion to 3.8 \%. This means that incoming atom kinetic 
energy assists growth in improving film density \cite{mull86} by redistributing 
atoms in metastable sites created by atomic roughness of the substrate. 
\section{Conclusions}
Molecular Dynamics simulations were conducted to study  cluster growth on 
atomically ordered, disordered and defect ordered surfaces up to 2 monolayers 
(3000 atoms). We use rigid lattice approximation and quenched MD which is
reasonable for deposition onto graphite surfaces.  Cluster growth follows
power law: $\bar{d} \propto t^{\alpha z}$,
with $\alpha z$ = 0.75 - 0.87. Influence of the type of disorder was 
investigated for different atom kinetic energies. The following trends are thus
observed. It was shown that roughness
can reduce sticking coefficients. Atomically
flat cluster are obtained on rough surface at room temperature while clusters
grown on atomically flat surface are rough. 
Roughness is also responsible for increasing low coverage cluster density and lattice
expansion. Increasing kinetic energy of incoming atom tends to minimize these 
roughness effects. But localized defect play a role that is not smoothed by
increasing kinetic energy: cluster growth is always disturbed by such a defect.
\section*{Acknowledgements}
This work was conducted using parallel computing facilities of CRIHAN under
 project 2000011 and partially supported by Contrat de Plan Etat/R\'egion 
 (CPER, fiche 15). M. C. Desjonqu\`eres, D. Spanjaard, G. Tr\'eglia , 
 C. Henry are gratefuly acknowledged for highlighting discusions and 
 encouragements. Special thanks to A.-L. Thomann, H. Estrade-Szwarckopf, B.
 Rousseau, C. Andreazza-Vignolle and P. Andreazza for our related experimental
 investigations.

\end{document}